# From the Standard Model to Dark Matter[*]

Frank Wilczek[†]

*School of Natural Sciences*
*Institute for Advanced Study*
*Olden Lane*
*Princeton, N.J. 08540*

## ABSTRACT

The standard model of particle physics is marvelously successful. However, it is obviously not a complete or final theory. I shall argue here that the structure of the standard model gives some quite concrete, compelling hints regarding what lies beyond. Taking these hints seriously, one is led to predict the existence of new types of very weakly interacting matter, stable on cosmological time scales and produced with cosmologically interesting densities–that is, "dark matter".

---

[*] Invited talk at 5th Annual October Maryland Astrophysics Conference
[†] Research supported in part by DOE grant DE-FG02-90ER40542. WILCZEK@IASSNS.BITNET

I have been asked to discuss particle physics candidates for dark matter. There are many ways one could go with such an assignment. I have made a very particular choice [1]. I will discuss the converging lines of thought and evidence leading from the consolidation and recent precision testing of the standard model to compelling ideas about unification of forces and the possibility of low-energy supersymmetry. These ideas produce as an important bonus a very attractive dark-matter candidate, the lightest supersymmetric particle – specified more precisely, using a concept to be defined below, as the lightest particle with odd R-parity.

Particle physics provides several other interesting and attractive dark matter candidates, notably including axions and massive neutrinos. It is entirely possible that one or both of these species provides a significant component of the mass density of the Universe. It is important vigorously to pursue experimental programs to detect each of them. Given the available time I had to make a choice, however. One pedagogical argument for presenting the supersymmetric option to an audience outside particle physics is that it is organically linked to the absolutely central theoretical ideas for going beyond the standard model, so I'll be able to lead into supersymmetric dark matter through presentation of these central ideas. Axions and neutrino masses are tied to important, but more peripheral ideas.

**Critique of the Standard Model**

The standard model of particle physics is based upon the gauge groups SU(3)×SU(2)×U(1) of strong, electromagnetic and weak interactions acting on the quark and lepton multiplets as shown in Figure 1.

In this Figure I have depicted only one family (u,d,e,$\nu_e$) of quarks and leptons; in reality there seem to be three families which are mere copies of one another as far as their interactions with the gauge bosons are concerned, but differ in mass. Actually in the Figure I have ignored masses altogether, and allowed myself the convenient fiction of pretending that the quarks and leptons have a definite chirality – right- or left-handed – as they would if they were massless. (The more precise statement, valid when masses are included, is that the gauge bosons couple to currents of definite chirality.) The chirality is indicated by a subscript R or L. Finally the little number beside each multiplet is its assignment under the U(1) of hypercharge, which is the average of the electric charge of the multiplet. (The physical photon is a linear combination of the diagonal generator of SU(2) and the hypercharge gauge bosons. The physical Z boson is the orthogonal combination.)

$$\begin{array}{ccccc} \text{SU(3)} & \times & \text{SU(2)} & \times & \text{U(1)} \\ \text{8 gluons} & & W^{\pm}, Z & & \gamma \end{array}$$

$$\underbrace{\qquad\qquad}_{\text{mixed}}$$

$$\text{SU(3)} \longleftrightarrow$$

$$\text{SU(2)} \updownarrow \begin{pmatrix} u_L^r & u_L^w & u_L^b \\ d_L^r & d_L^w & d_L^b \end{pmatrix} \frac{1}{6} \quad (u_R^r \quad u_R^w \quad u_R^b)\frac{2}{3}$$
$$\qquad\qquad\qquad\qquad\qquad\qquad\qquad (d_R^r \quad d_R^w \quad d_R^b) -\frac{1}{3}$$

$$\begin{pmatrix} \nu_L \\ e_L \end{pmatrix} -\frac{1}{2} \quad e_R - 1$$

**FIGURE 1**

*Figure 1 - The gauge groups of the standard model, and the fermion multiplets with their hypercharges.*

Figure 1, properly understood – that is, the standard model – describes a tremendous amount of physics. The strong interactions responsible for the structure of nucleons and nuclei, and for most of what happens in high energy collisions; the weak interactions responsible for nuclear transmutations; and the electromagnetic interactions responsible in Dirac's phrase for "all of chemistry and most of physics" are all there, described by mathematically precise and indeed rather similar theories of vector gauge particles interaction with spin-$\frac{1}{2}$ fermions. The standard model provides a remarkably compact description of all this. It is also a remarkably successful description, with its fundamentals having now been vigorously and rigorously tested in many experiments, especially at LEP. Precise quantitative comparisons between theory and experiment are nothing new for QED and the weak interactions, but if you haven't been paying attention you may not be aware that the situation for QCD has improved dramatically in the last few years [2]. For example phenomenologists now debate over the third decimal place in the strong coupling constant, experiments are now routinely sensitive to two-loop and even three-loop QCD effects, and recent lattice gauge simulations are achieving 10% or better accuracy in the spectrum both for heavy quark and for light quark systems [3].

While little doubt can remain that the standard model is essentially correct, a glance at Figure 1 is enough to reveal that it is not a complete or final theory. The fermions fall into apart into five lopsided pieces with peculiar hypercharge assignments; this pattern needs to be explained. Also the separate gauge theories, which as I mentioned are mathematically similar, are fairly begging to be unified. Let me elaborate a bit on this. The SU(3) of strong interactions is, roughly speaking, an extension of QED to three new types of charges, which in the QCD context are called colors (say red, white, and blue). QCD contains eight different gauge boson, or color gluons. There are six possible gauge bosons which transform one unit of any color charge into one unit of any other, and two photon-like gauge bosons that sense the colors. An important subtlety which emerges simply from the mathematics and which will play an important role in our further considerations is that there are two rather than three color-sensing gauge bosons. This is because the linear combination which couples to all three color charges equally is not part of SU(3). Similarly the SU(2) of weak interactions is the theory of two colors (say green and purple) and features three gauge bosons: the weak color changing ones, which we call $W^+$, $W^-$, and the weak color-sensing one that mixes with the U(1) hypercharge boson to yield Z and the photon $\gamma$.

### Unification: quantum numbers

Given that the strong interactions are governed by transformations among three colors, and the weak by transformations between two others, what could be more natural than to embed both theories into a larger theory of transformations among all five colors? This idea has the additional attraction that an extra U(1) symmetry commuting with the strong SU(3) and weak SU(2) symmetries automatically appears, which we can attempt to identify with the remaining gauge symmetry of the standard model, that is hypercharge. For while in the separate SU(3) and SU(2) theories we must throw out the two gauge bosons which couple respectively to the color combinations R+W+B and G+P, in the SU(5) theory we only project out R+W+B+G+P, while the orthogonal combination (R+W+B)-$\frac{3}{2}$(G+P) remains.

Georgi and Glashow [4] originated this line of thought, and showed how it could be used to bring some order to the quark and lepton sector, and in particular to supply a satisfying explanation of the weird hypercharge assignments in the standard model. As shown in Figure 2, the five scattered SU(3)×SU(2)×U(1) multiplets get organized into just two representations of SU(5).

In making this unification it is necessary to allow transformations between (what were previously thought to be) particles and antiparticles of the same chirality, and also between quarks and leptons. It is convenient to work with left-handed fields only; since the conjugate of a right-handed field is left-handed, we don't lose track of anything by doing so, once we disabuse ourselves of the idea that a given field is intrinsically either genuine or "anti".

As shown in Figure 2, there is one group of ten left-handed fermions that have all possible combinations of one unit of each of two different colors, and another group of five left-handed fermions that each carry just one negative unit of some color. (These are the ten-dimensional antisymmetric tensor and the complex conjugate of the five-dimensional vector representation, commonly referred to as the "five-bar".) What is important for you to take away from this discussion is not so much the precise details of the scheme, but the idea that *the structure of the standard model, with the particle assignments gleaned from decades of experimental effort and theoretical interpretation, is perfectly reproduced by a simple abstract set* *of rules for manipulating symmetrical symbols.* Thus for example the object RB in this Figure has just the strong, electromagnetic, and weak interactions we expect of the complex conjugate of the right-handed up-quark, without our having to instruct the theory further. If you've never done it I heartily recommend to you the simple exercise of working out the hypercharges of the objects in Figure 2 and checking against what you need in the standard model – after doing it, you'll find it's impossible ever to look at the standard model in quite the same way again.

SU(5): 5 colors RWBGP

<u>10</u>: 2 different color labels (antisymmetric tensor)

$$u_L : \quad RP, \quad WP, \quad BP$$
$$d_L : \quad RG, \quad WG, \quad BG$$
$$u_L^c : \quad RW, \quad WB, \quad BR$$
$$\quad\quad\quad (\bar{B}) \quad (\bar{R}) \quad (\bar{W})$$
$$e_L^c : \quad GP$$
$$\quad\quad ( \; )$$

$$\begin{pmatrix} 0 & u^c & u^c & u & d \\ & 0 & u^c & u & d \\ & & 0 & u & d \\ * & & & 0 & e \\ & & & & 0 \end{pmatrix}$$

<u>5̄</u>: 1 anticolor label

$$d_L^c : \quad \bar{R}, \quad \bar{W}, \quad \bar{B}$$
$$e_L : \quad \bar{P} \quad\quad (d^c \quad d^c \quad d^c \quad e \quad \nu)$$
$$\nu_L : \quad \bar{G}$$

$$\boxed{Y = -\tfrac{1}{3}(R+W+B) + \tfrac{1}{2}(G+P)}$$

**FIGURE 2**

*Figure 2 - Unification of fermions in SU(5).*

There is a beautiful extension of SU(5) to a slightly larger group, SO(10), which permits one to unite all the fermions of a family into a single multiplet. The relevant representations for the fermions is a 16-dimensional spinor representation. Some of its features are depicted in Figure 3, as I shall now explain.

$$(\pm \pm \pm \pm \pm) \; : \; \underline{\text{even}} \; \# \; \text{of} \; -$$

$$10 : \begin{matrix} (++-|+-) & 6 & (u_L, d_L) \\ (+--|++) & 3 & u_L^c \\ (+++|--) & 1 & e_L^c \end{matrix}$$

$$\bar{5} : \begin{matrix} (+--|--) & \bar{3} & d_L^c \\ (---|+-) & \bar{2} & (e_L, \nu_L) \end{matrix}$$

$$1 : (+++|++) \quad 1 \quad N_R$$

**FIGURE 3**

*Figure 3 - Unification of fermions in SO(10). The rule is that all possible combinations of 5 + and - signs occur, subject to the constraint that the total number of - signs is even. The SU(5) gauge bosons within SO(10) do not change the numbers of signs, and one sees the SU(5) multiplets emerging. However there are additional transformations in SO(10) but not in SU(5), which allow any fermion to be transformed into any other. Permutations of signs within the first three slots or within the last three slots are not indicated. The numbers in the left-hand column indicates the SU(5) quantum multiplets – to be compared with Figure 2; the numbers in the third column indicates the multiplicity of standard model multiplets – to be compared with Figure 1.*

Spinor representations are most easily constructed iteratively. To construct spinors for rotations in n-dimensional space, one needs an algebra of $\gamma$-matrices obeying the anticommutation relations

$$\{\gamma_i^{(n)}, \gamma_j^{(n)}\} = 2\delta_{ij} \qquad (1)$$

for $i, j \leq n$. For $n = 2$ one can use the first two Pauli matrices: $\gamma_1^{(2)} = \tau_1 = \begin{pmatrix} 0 & 1 \\ 1 & 0 \end{pmatrix}$, $\gamma_2^{(2)} = \tau_2 = \begin{pmatrix} 0 & -i \\ i & 0 \end{pmatrix}$. Then to move from $n - 2$ to $n$ dimensions one uses the set

$$\gamma_i^{(n)} = \gamma_i^{(n-2)} \otimes \tau_3 \qquad (2)$$

for $i \leq n - 2$ and

$$\begin{aligned} \gamma_{n-1}^{(n)} &= 1 \otimes \tau_1 \\ \gamma_n^{(n)} &= 1 \otimes \tau_2 \end{aligned} \qquad (3)$$

where of course in (3) 1 denotes the $2^{\frac{n}{2}} \times 2^{\frac{n}{2}}$ dimensional unit matrix. One can easily unfold the $\gamma^{(n)}$ into n-fold tensor products, or alternatively write down explicit ordinary matrix representations. Thus for example

$$\gamma_1^{(4)} = \begin{pmatrix} 0 & 1 & 0 & 0 \\ 1 & 0 & 0 & 0 \\ 0 & 0 & 0 & -1 \\ 0 & 0 & -1 & 0 \end{pmatrix} \quad \gamma_2^{(4)} = \begin{pmatrix} 0 & -i & 0 & 0 \\ i & 0 & 0 & 0 \\ 0 & 0 & 0 & i \\ 0 & 0 & -i & 0 \end{pmatrix}$$

$$\gamma_3^{(4)} = \begin{pmatrix} 0 & 0 & 1 & 0 \\ 0 & 0 & 0 & 1 \\ 1 & 0 & 0 & 0 \\ 0 & 1 & 0 & 0 \end{pmatrix} \quad \gamma_4^{(4)} = \begin{pmatrix} 0 & 0 & -i & 0 \\ 0 & 0 & 0 & -i \\ i & 0 & 0 & 0 \\ 0 & i & 0 & 0 \end{pmatrix} \qquad (4)$$

and

$$\gamma_1^{(6)} = \begin{pmatrix} 0 & 1 & 0 & 0 & 0 & 0 & 0 & 0 \\ 1 & 0 & 0 & 0 & 0 & 0 & 0 & 0 \\ 0 & 0 & 0 & -1 & 0 & 0 & 0 & 0 \\ 0 & 0 & -1 & 0 & 0 & 0 & 0 & 0 \\ 0 & 0 & 0 & 0 & 0 & -1 & 0 & 0 \\ 0 & 0 & 0 & 0 & -1 & 0 & 0 & 0 \\ 0 & 0 & 0 & 0 & 0 & 0 & 0 & 1 \\ 0 & 0 & 0 & 0 & 0 & 0 & 1 & 0 \end{pmatrix} \quad \gamma_2^{(6)} = \begin{pmatrix} 0 & -i & 0 & 0 & 0 & 0 & 0 & 0 \\ i & 0 & 0 & 0 & 0 & 0 & 0 & 0 \\ 0 & 0 & 0 & i & 0 & 0 & 0 & 0 \\ 0 & 0 & -i & 0 & 0 & 0 & 0 & 0 \\ 0 & 0 & 0 & 0 & 0 & i & 0 & 0 \\ 0 & 0 & 0 & 0 & -i & 0 & 0 & 0 \\ 0 & 0 & 0 & 0 & 0 & 0 & 0 & -i \\ 0 & 0 & 0 & 0 & 0 & 0 & i & 0 \end{pmatrix}$$

$$\gamma_3^{(6)} = \begin{pmatrix} 0 & 0 & 1 & 0 & 0 & 0 & 0 & 0 \\ 0 & 0 & 0 & 1 & 0 & 0 & 0 & 0 \\ 1 & 0 & 0 & 0 & 0 & 0 & 0 & 0 \\ 0 & 1 & 0 & 0 & 0 & 0 & 0 & 0 \\ 0 & 0 & 0 & 0 & 0 & 0 & -1 & 0 \\ 0 & 0 & 0 & 0 & 0 & 0 & 0 & -1 \\ 0 & 0 & 0 & 0 & -1 & 0 & 0 & 0 \\ 0 & 0 & 0 & 0 & 0 & -1 & 0 & 0 \end{pmatrix} \quad \gamma_4^{(6)} = \begin{pmatrix} 0 & 0 & -i & 0 & 0 & 0 & 0 & 0 \\ 0 & 0 & 0 & -i & 0 & 0 & 0 & 0 \\ i & 0 & 0 & 0 & 0 & 0 & 0 & 0 \\ 0 & i & 0 & 0 & 0 & 0 & 0 & 0 \\ 0 & 0 & 0 & 0 & 0 & 0 & i & 0 \\ 0 & 0 & 0 & 0 & 0 & 0 & 0 & i \\ 0 & 0 & 0 & 0 & -i & 0 & 0 & 0 \\ 0 & 0 & 0 & 0 & 0 & -i & 0 & 0 \end{pmatrix}$$

$$\gamma_5^{(6)} = \begin{pmatrix} 0 & 0 & 0 & 0 & 1 & 0 & 0 & 0 \\ 0 & 0 & 0 & 0 & 0 & 1 & 0 & 0 \\ 0 & 0 & 0 & 0 & 0 & 0 & 1 & 0 \\ 0 & 0 & 0 & 0 & 0 & 0 & 0 & 1 \\ 1 & 0 & 0 & 0 & 0 & 0 & 0 & 0 \\ 0 & 1 & 0 & 0 & 0 & 0 & 0 & 0 \\ 0 & 0 & 1 & 0 & 0 & 0 & 0 & 0 \\ 0 & 0 & 0 & 1 & 0 & 0 & 0 & 0 \end{pmatrix} \quad \gamma_6^{(6)} = \begin{pmatrix} 0 & 0 & 0 & 0 & -i & 0 & 0 & 0 \\ 0 & 0 & 0 & 0 & 0 & -i & 0 & 0 \\ 0 & 0 & 0 & 0 & 0 & 0 & -i & 0 \\ 0 & 0 & 0 & 0 & 0 & 0 & 0 & -i \\ i & 0 & 0 & 0 & 0 & 0 & 0 & 0 \\ 0 & i & 0 & 0 & 0 & 0 & 0 & 0 \\ 0 & 0 & i & 0 & 0 & 0 & 0 & 0 \\ 0 & 0 & 0 & i & 0 & 0 & 0 & 0 \end{pmatrix}$$
(5)

Clearly the dimension of the $\gamma^{(n)}$ matrices is doubling each time $n$ increases by two. (We will not need to be concerned with odd $n$ here, but of course if $n$ is odd we can satisfy (1) by using the first $n$ $\gamma^{(n+1)}$s.)

Now it is very easy to verify from (1) that the matrices

$$\sigma_{ij}^{(n)} \equiv \frac{-i}{4}[\gamma_i^{(n)}, \gamma_j^{(n)}]$$

satisfy the commutation relations

$$i[\sigma_{ij}^{(n)}, \sigma_{kl}^{(n)}] = \delta_{ik}\sigma_{jl}^{(n)} - \delta_{il}\sigma_{jk}^{(n)} . \tag{6}$$

These are the same commutation relations one has for the infinitesimal rotations around the $ij$ and $kl$ axes. Thus we have represented the Lie algebra of rotations in n dimensions by a set of $2^{\frac{n}{2}} \times 2^{\frac{n}{2}}$ matrices; that is, we have a $2^{\frac{n}{2}}$ dimensional representation of $SO(n)$. This representation is not quite irreducible, however. One easily verifies that the product

$$\Gamma^{(n)} = (-i)^{\frac{n}{2}} \gamma_1^{(n)} \gamma_2^{(n)} \cdots \gamma_n^{(n)} \tag{7}$$

anticommutes with all the $\gamma_i^{(n)}$, and therefore commutes with all the $\sigma_{ij}^{(n)}$. Indeed, in the specific representation we have chosen $\Gamma^{(n)}$ unfolds into the n-fold tensor product

$$\Gamma^{(n)} = \tau_3 \otimes \tau_3 \cdots \otimes \tau_3 . \tag{8}$$

We also have

$$(\Gamma^{(n)})^2 = 1 . \tag{9}$$

Thus by projecting onto the eigenspaces of $\Gamma^{(n)}$, using the projection operators $(1 \pm \Gamma^{(n)})/2$, we find two representations of dimension $2^{\frac{n}{2}-1}$. These representations turn out to be irreducible. The representation we shall be most interested in is the 16 dimensional representation of SO(10).

Now at last we are in position to make contact with the notations of Figure 3. One can label states in the spinor representation, as we have constructed it, very conveniently in terms of the eigenvalues of the operators

$$2\sigma_{12}^{(10)} = \tau_3 \otimes 1 \otimes 1 \otimes 1 \otimes 1$$
$$2\sigma_{34}^{(10)} = 1 \otimes \tau_3 \otimes 1 \otimes 1 \otimes 1 \quad (10)$$
$$\cdots .$$

Indeed, these form a complete set of commuting observables, so their simultaneous eigenvalues can be used to label the states. In this way, we see that each state in the spinor representation, as we have constructed it, corresponds to an ordered choice of 5 $\pm$ signs. Projection with $(1 + \Gamma^{(10)})/2$ corresponds to imposing the constraint that the product of these signs is positive, *i.e.* that there should be an even number of - signs. Thus we arrive at the 16 states depicted in Figure 3.

For the application to physics it is important to identify explicitly the quantum numbers of the standard model within the abstract realization of the postulated unified gauge symmetry group. How does the symmetry of the standard model sit within $SO(10)$, and how do particles in the spinor 16 representation transform under it? Well, the group $SU(n)$ forms a subgroup of $SO(2n)$ in a very canonical way. Indeed, $SU(n)$ is the group that preserves a Hermitean inner product between vectors in a n-complex dimensional vector space. But the real part of this inner product is just the ordinary real inner product of the 2n-real dimensional vectors formed from the real and imaginary parts of n-complex dimensional vectors, so $SU(n)$ is the subgroup of $SO(2n)$, which leaves the imaginary part as well as the real part of the inner product invariant.

Thus the $SU(3) \times SU(2)$ part of the standard model is easy to locate within $SO(6) \times SO(4)$ inside $SO(10)$. If we declare that the components 2,4,6,8,10 are the imaginary parts of the complex vectors whose real parts are the components 1,3,5,7,9 respectively, then $SU(3)$ will naturally act on the first six components – that is, it will consist of suitable combinations of the $\sigma_{ij}^{(10)}$ with $i,j \leq 6$; and $SU(2)$ will consist of suitable combinations of the $\sigma_{ij}^{(10)}$ with $7 \leq i,j$.

With this way of organizing things, it becomes easy to identify the physical meaning of the operators in (10) and the $\pm$ signs used to label the states. Indeed, each operator in (10) becomes the generator of an infinitesimal *phase rotation* of one of the five complex components of vectors in $SU(5)$. The first three of these correspond to generators of QCD color charges, say red, white, and blue. It is natural to call the final two the generators weak color charges, say green and purple. The generators

$$\Sigma_{\text{QCD}} = \sigma_{12}^{(10)} + \sigma_{34}^{(10)} + \sigma_{56}^{(10)} \qquad (11)$$

and

$$\Sigma_{\text{weak}} = \sigma_{78}^{(10)} + \sigma_{9,10}^{(10)} \qquad (12)$$

represent phase rotations that commute with $SU(3) \times SU(2)$. Indeed $\Sigma_{\text{QCD}} + \Sigma_{\text{weak}}$ generates a common phase rotation of all 5 complex vector components. It is not properly in $SU(5)$ at all; that's the difference between $SU(5)$ and just $U(5)$. The hypercharge generator in $SU(5)$ is proportional to the 'traceless' combination $2\Sigma_{\text{QCD}} - 3\Sigma_{\text{weak}}$.

By the way, $U(5)$ can be located within $SO(10)$ by the condition that $U(5)$ generators are the combinations of $SO(10)$ generators that commute with $\Sigma_{\text{QCD}} + \Sigma_{\text{weak}} \equiv J$. Indeed $J$ implements, as we have seen, an overall phase rotation. The condition that a linear transformation which already leaves the real part of the inner product invariant should also leave the imaginary part invariant is exactly that it respect such a phase rotation.

All the concepts used in constructing Figure 3 have now been spelled out, and at this point it ought to be a pleasant exercise for you to verify that each "particle" constructed in this abstract mathematical way has just the quantum numbers to be identified with one of the fundamental fermions (with one interesting exception, as we shall discuss immediately below). In this accounting, one represents each fermion using a left-handed chiral field, by taking charge conjugates if necessary. Thus if one wishes to find the right-handed up quark, for instance, one should look for the left-handed anti-up antiquark. For an example, consider $(+ + - + -)$. It forms part of a QCD triplet together with $(+ - + + -)$ and $(- + + + -)$, and part of a weak doublet together with $(+ + - - +)$. $(+ + + - -)$, on the contrary, is a singlet both for QCD and for weak interactions. The hypercharges of these two objects are in the ratio 2:12. These are exactly the quantum numbers one wants for $(+ + - + -)$ to be a component of a left-handed quark field, and $(+ + + - -)$ the charge conjugate of $e_R$, the right-handed component of the electron.

Proceeding along these lines, one finds an uncanny fit between the abstract quantum numbers of $SO(10)$ spinors and the ones observed for particles in the real world: the charge spectrum, and all the strong, electromagnetic, and weak interactions of the standard model, are incorporated. There is also one extra state in the model, however, namely $(+ + + + +)$. This state is a singlet under all the interactions of the standard model. Thus it is not surprising that the particle corresponding to this state would escape easy detection, even were it to exist. One can make very good use of this state in constructing models of massive neutrinos, but I promised not to get into that subject.

What I have shown you here is very old and standard mathematics. Whatever small novelty there is, is in the packaging. That said, I must admit that I find this way of presenting things, culminating in the labeling of fermion states by ordered bits as displayed in Figure 3, very appealing and seductive. I beg indulgence to mention two sorts of fantasies it suggests. First, of course, one might speculate that there are additional colors, thus being led to $SO(10 + x)$ theories. This large symmetry group can be put to use in attempting to address the question why there are multiple families of fermions with identical $SO(10)$ (or at least standard model) quantum numbers. It is a remarkable fact, quite transparent in our construction of the spinor representation, that a spinor of $SO(10 + n)$ breaks up into several spinors (and an equal number of antispinors) under $SO(10)$, with no other representations appearing. For some adventures in trying to exploit these ideas see [6]. Alternatively, one might speculate that the representation of particles as bit-structures is a profound feature of the physical world, conceivably more fundamental than any particular gauge group or even than gauge theory altogether. A relatively modest (but conversely, relatively concrete) idea in this vein is that the different particles should be described as some sort of soliton or magnetic monopole, and the various signs indicate occupation (or not) of a set of zero modes.

### Unification: coupling values

We have seen that simple unification schemes are successful at the level of *classification*; but new questions arise when we consider the dynamics which underlies them.

Part of the power of gauge symmetry is that it fully dictates the interactions of the gauge bosons, once an overall coupling constant is specified. Thus if SU(5) or some higher symmetry were exact, then the fundamental strengths of the different color-changing interactions would have to be equal, as would the (properly normalized) hypercharge coupling strength. In reality the coupling strengths of the gauge bosons in SU(3)×SU(2)×U(1) are not observed to be equal, but rather follow the pattern $g_3 \gg g_2 > g_1$.

Fortunately, experience with QCD emphasizes that couplings "run". The physical mechanism of this effect is that in quantum field theory the vacuum must be regarded as a polarizable medium, since virtual particle-anti-particle pairs can screen charge. Thus one might expect that effective charges measured at shorter distances, or equivalently at larger energy-momentum or mass scales, could be different from what they appear at longer distances. If one had only screening then the effective couplings would grow at shorter distances, as one penetrated deeper insider the screening cloud. However it is a famous fact [7] that due to paramagnetic spin-spin attraction of like charge vector gluons [8], these particles tend to *antiscreen* color charge, thus giving rise to the opposite effect – asymptotic freedom – that the effective coupling tends to shrink at short distances. This effect is the basis of all perturbative QCD phenomenology, which is a vast and vastly successful enterprise. For our present purpose of understanding the disparity of the observed couplings, it is just what the doctor ordered. As was first pointed out by Georgi,

Quinn, and Weinberg [9], if a gauge symmetry such as SU(5) is spontaneously broken at some very short distance then we should not expect that the effective couplings probed at much larger distances, such as are actually measured at practical accelerators, will be equal. Rather they will all have have been affected to a greater or lesser extent by vacuum screening and anti-screening, starting from a common value at the unification scale but then diverging from one another. The pattern $g_3 \gg g_2 > g_1$ is just what one should expect, since the antiscreening or asymptotic freedom effect is more pronounced for larger gauge groups, which have more types of virtual gluons.

**FIGURE 4**

*Figure 4 - The failure of the running couplings, normalized according to SU(5) and extrapolated taking into account only the virtual exchange of the "known" particles of the standard model (including the top quark and Higgs boson) to meet. Note that only with quite recent experiments, which greatly improved the precision of the determination of low-energy couplings, did the discrepancy become significant.*

The marvelous thing is that the running of the couplings gives us a truly quantitative handle on the ideas of unification, for the following reason. To fix the relevant aspects of unification, one basically needs only to fix two parameters: the scale at which the couplings unite, which is essentially the scale at which the unified symmetry breaks; and their value when then unite. Given these, one calculates three outputs: the three *a priori* independent couplings for the gauge groups in SU(3)×SU(2)×U(1). Thus the framework is eminently falsifiable. The miraculous thing is, how close it comes to working (Figure 4).

The unification of couplings occurs at a very large mass scale, $M_{\rm un.} \sim 10^{15}$ Gev. In the simplest version, this is the magnitude of the scalar field vacuum expectation value that spontaneously breaks SU(5) down to the standard model symmetry SU(3)×SU(2)×U(1), and is analogous to the scale $v \approx 250$ Gev for electroweak symmetry breaking. The largeness of this large scale mass scale is important in several ways.

• It explains why the exchange of gauge bosons that are in SU(5) but not in SU(3)×SU(2)×U(1), which reshuffles strong into weak colors and generically violates the conservation of baryon number, does not lead to a catastrophically quick decay of nucleons. The rate of decay goes as the inverse fourth power of the mass of the exchanged gauge particle, so the baryon-number violating processes are predicted to be far slower than ordinary weak processes, as they had better be.

• $M_{\rm un.}$ is significantly smaller than the Planck scale $M_{\rm Planck} \sim 10^{19}$ Gev at which exchange of gravitons competes quantitatively with the other interactions, but not ridiculously so. This indicates that while the unification of couplings calculation itself is probably safe from gravitational corrections, the unavoidable logical next step in unification must be to bring gravity into the mix.

• Finally one must ask how the tiny ratio of symmetry-breaking mass scales $v/M_{\rm un.} \sim 10^{-13}$ required arises dynamically, and whether it is stable. This is the so-called gauge hierarchy problem, which we shall discuss in a more concrete form a little later.

The success of the GQW calculation in explaining the observed hierarchy $g_3 \gg$

$g_2 > g_1$ of couplings and the approximate stability of the proton is quite striking. In performing it, we assumed that the known and confidently expected particles of the standard model exhaust the spectrum up to the unification scale, and that the rules of quantum field theory could be extrapolated without alteration up to this mass scale – thirteen orders of magnitude beyond the domain they were designed to describe. It is a triumph for minimalism, both existential and conceptual.

However, on further examination it is not quite good enough. Accurate modern measurements of the couplings show a small but definite discrepancy between the couplings, as appears in Figure 4. And heroic dedicated experiments to search for proton decay did not find it [10]; they currently exclude the minimal SU(5) prediction $\tau_p \sim 10^{31}$ yrs. by about two orders of magnitude.

Given the magnitude of the extrapolation involved, perhaps we should not have hoped for more. There are several perfectly plausible bits of physics that could upset the calculation, such as the existence of particles with masses much higher than the electroweak but much smaller than the unification scale. As virtual particles these would affect the running of the couplings, and yet one certainly cannot exclude their existence on direct experimental grounds. If we just add particles in some haphazard way things will only get worse: minimal SU(5) nearly works, so the generic perturbation from it will be deleterious. This is a major difficulty for so-called technicolor models, which postulate many new particles in complex patterns. Even if some *ad hoc* prescription could be made to work, that would be a disappointing outcome from what appeared to be one of our most precious, elegantly straightforward clues regarding physics well beyond the standard model.

**Virtual supersymmetry?**

Fortunately, there is a theoretical idea which is attractive in many other ways, and seems to point a way out from this impasse. That is the idea of supersymmetry [11]. Supersymmetry is a symmetry that extends the Poincare symmetry of special relativity (there is also a general relativistic version). In a supersymmetric theory one has not only transformations among particle states with different energy-momentum but also between particle states of different *spin*. Thus spin 0 particles can be put in multiplets together with spin $\frac{1}{2}$ particles, or spin $\frac{1}{2}$ with spin 1, and so forth.

Supersymmetry is certainly not a symmetry in nature: for example, there is certainly no bosonic particle with the mass and charge of the electron. More generally if one defines the $R$-parity quantum number

$$R \equiv (-)^{3B+L+2S} ,$$

which should be accurate to the extent that baryon and lepton number are conserved, then one finds that all currently known particles are $R$ even whereas their supersymmetric partners would be $R$ odd. Nevertheless there are many reasons to be interested in supersymmetry, of which I shall mention three.

• You will notice that we have made progress in uniting the gauge bosons with each other, and the various quarks and leptons with each other, but not the gauge bosons with the quarks and leptons. It takes supersymmetry – perhaps spontaneously broken – to make this feasible.

• Supersymmetry was invented in the context of string theory, and seems to be necessary for constructing consistent string theories containing gravity (critical string theories) that are at all realistic.

**FIGURE 5**

*Figure 5 - A typical quadratically divergent contribution to the (mass)² of the Higgs boson, and the supersymmetric contribution which, as long as supersymmetry is not too badly broken, will largely cancel it.*

• Most important for our purposes, supersymmetry can help us to understand the vast disparity between weak and unified symmetry breaking scales mentioned above. This disparity is known as the gauge hierarchy problem. It actually raises several distinct problems, including the following. In calculating radiative corrections to the (mass)² of the Higgs particle from diagrams of the type shown in Figure 5 one finds an infinite, and also large, contribution. By this I mean that the divergence is quadratic in the ultraviolet cutoff. No ordinary symmetry will make its coefficient vanish. If we imagine that the unification scale provides the cutoff, we find that the radiative correction to the (mass)² is much larger than the final value we want. (If the Higgs field were composite, with a soft form factor, this problem might be ameliorated. Following that road leads to technicolor, which as mentioned before seems to lead us far away from our best source of inspiration.) As a formal matter one can simply cancel the radiative correction against a large bare contribution of the opposite sign, but in the absence of some deeper motivating principle this seems to be a horribly ugly procedure. Now in a supersymmetric theory for any set of virtual particles circulating in the loop there will also be another graph with their supersymmetric partners circulating. If the partners were accurately degenerate, the contributions would cancel. Otherwise, the threatened quadratic divergence will be cut off only at virtual momenta such that the difference in (mass)$^2$ between the virtual particle and its supersymmetric partner is negligible. Thus we will be assured adequate cancelation if and only if supersymmetric partners are not too far split in mass – in the present context, if the splitting is not much greater than the weak scale. This is (a crude version of) the most important *quantitative* argument which suggests the relevance of "low-energy" supersymmetry.

The effect of low-energy supersymmetry on the running of the couplings was first considered long ago [12], well before the crisis described at the end of the previous section was evident. One might fear that such a huge expansion of the theory, which essentially doubles the spectrum, would utterly destroy the approximate success of the minimal SU(5) calculation. This is not true, however. To a first approximation since supersymmetry is a space-time rather than an internal symmetry it does not affect the group-theoretic structure of the calculation. Thus to a first approximation the absolute rate at which the couplings run with momentum is affected, but not the relative rates. The main effect is that the supersymmetric partners of the color gluons, the gluinos, weaken the asymptotic freedom of the strong interaction. Thus they tend to make its effective coupling decrease and approach the others more slowly. Thus their merger requires a longer lever arm, and the scale at which the couplings meet increases by an order of magnitude or so, to about $10^{16}$ Gev. Also the common value of the effective couplings at unification is slightly larger than in conventional unification ($\frac{g_{un.}^2}{4\pi} \approx \frac{1}{25}$ *versus* $\frac{1}{40}$). This increase in unification scale significantly reduces the predicted rate for proton decay through exchange of the dangerous color-changing gauge bosons, so that it no longer conflicts with existing experimental limits.

Upon more careful examination there is another effect of low-energy supersymmetry on the running of the couplings, which although quantitatively small has become of prime interest. There is an important exception to the general rule that adding supersymmetric partners does not immediately (at the one loop level) affect the relative rates at which the couplings run. This rule works for particles that come in complete SU(5) multiplets, such as the quarks and leptons (which, since they don't upset the full SU(5) symmetry, have basically no effect) or for the supersymmetric partners of the gauge bosons, because they just renormalize the existing, dominant effect of the gauge bosons themselves. However there is one peculiar additional contribution, from the supersymmetric partner of the Higgs doublet. It affects only the weak SU(2) and hypercharge U(1) couplings. (On phenomenological grounds the SU(5) color triplet partner of the Higgs doublet must be extremely massive, so its virtual exchange is not important below the unification scale. *Why* that should be so, is another aspect of the hierarchy problem.) Moreover, for slightly technical reasons even in the minimal supersymmetric model it is necessary to have two different Higgs doublets with opposite hypercharges. The net affect of doubling the number of Higgs fields and including their supersymmetric partners is a sixfold enhancement of the asymmetric Higgs field contribution to the running of weak and hypercharge couplings. This causes a small, accurately calculable change in the calculation. From Figure 6 you see that it is a most welcome one. Indeed, in the minimal implementation of supersymmetric unification, it puts the running of couplings calculation right back on the money [13].

Since the running of the couplings with scale is logarithmic the unification of couplings calculation is not terribly sensitive to the exact scale at which supersymmetry is broken, say between 100 Gev and 10 Tev. There have been attempts to push the calculation further, in order to address this question of the supersymmetry breaking scale, but they are controversial. It is not obvious to me that such calculations will ever achieve the resolution of interest.

For example, comparable uncertainties arise from the splittings among the very large number of particles with masses of order the unification scale, whose theory is poorly developed and unreliable.

**FIGURE 6**

*Figure 6 - When the exchange of the virtual particles necessary to implement low-energy supersymmetry, a calculation along the lines of Figure 4 comes into adequate agreement with experiment.*

In any case, if we are not too greedy the main points still shine through:

- If supersymmetry is to fulfill its destiny of elucidating the hierarchy problem in any straightforward way, then the supersymmetric partners of the known particles cannot be much heavier than the SU(2)×U(1) electroweak breaking scale, *i.e.* they should not be beyond the expected reach of LHC.
- If we assume this to be the case then the meeting of the couplings takes place in the simplest minimal models of unification, without further assumption – a most remarkable and non-trivial fact.

To the extent R-parity is valid, the lightest R-odd particle is forbidden to decay. Since baryon number and lepton number, not to mention spin, are rather accurately conserved this particle has a chance to be extremely stable. Since it can annihilate in pairs, there is no possibility of an "intrinsic asymmetry" – thus the density of relic particles left over from the big bang can be calculated in a straightforward fashion, in the given a model of particle physics. It has been known for a long time that particles with annihilation cross-section of roughly weak interaction strength and masses of a few Gev. would be produced with cosmologically interesting densities$^\star$. Given this encouragement from big bang cosmology, in the context of the foregoing discussion, I think you will agree that the lightest R-odd particle is a most interesting candidate to provide cosmological dark matter.

In the conservative kinds of models that try to make minimal additions to the standard model, while incorporating the advantages for unification of couplings and stability against large radiative corrections that I mentioned before, the lightest R-odd particle usually turns out to be a linear combination of the supersymmetric partners of the neutral Higgs particles, and the photon and Z bosons. These various partners are called the higgsino, photino, and bino, respectively, and the combination of definite mass is called the neutralino. It is a spin-1/2 electrically neutral particle.

There is considerable uncertainty in the predictions for the mass and interaction properties of the lightest supersymmetric particle, since at the present stage of knowledge many parameters in the models must be taken as free variables. For example, in some ranges of parameters the lightest supersymmetric particle turns out to be charged. This is difficult to reconcile with the idea that it is cosmologically stable, since there are extremely powerful experimental constraints on such particles. Their cosmology is also problematic – they would presumably dissipate, fall into the disc, find one another and annihilate efficiently giving a high-energy photon background, ... . Dark matter really ought to be dark (or, more accurately,

---

$\star$ The principles for such calculations were first discussed, mainly in the context of searching for relic quarks, by Zeldovich. Another seminal contribution was a very clear and influential paper on heavy (that is, several Gev.) neutrinos by Lee and Weinberg [14].

transparent). In other ranges of parameters the lightest R-odd particle turns out to be the supersymmetric partner of the neutrino. This raises many interesting additional issues (including the possibility of an intrinsic asymmetry!), but resembles the neutralino scenario in broad outline.

There is a sizable literature devoted to discussions of the best methods for detecting supersymmetric particles, including excellent reviews [15, 16]. Both more-or-less conventional techniques of particle physics to produce and sense the superparticles at higher energy accelerators, and extraordinary techniques to sense the cosmological background itself have been contemplated. Supersymmetry also provides new mechanisms for CP violation, proton decay, and flavor-changing precesses that could come in at experimentally detectable levels.

I hope I've been able to convey to you a few core ideas for physics beyond the standard model that can be understood fairly simply and that appear likely to be of permanent value. They provide, in my opinion, very good specific reasons to be hopeful about the future of experimental particle physics, and related domains of cosmology, if we can summon up the national or international will to pursue it. One hears the distant rumbling of big game afoot.